\documentclass[twocolumn,aps,prl,showpacs,nofootinbib,floatfix]{revtex4}
\usepackage{amsmath,latexsym,times,graphicx}
\begin{document}
\title{Gravitational radiation reaction and inspiral waveforms
in the adiabatic limit}

\author{Scott A.\ Hughes$^{1,2}$, Steve Drasco$^3$, \'Eanna \'E.\
Flanagan$^{3}$ and Joel Franklin$^2$}
\affiliation{$^1$Department of Physics, MIT, 77 Massachusetts Ave.,
Cambridge, MA 02139}
\affiliation{$^2$Center for Space Research, MIT, 77 Massachusetts
Ave., Cambridge, MA 02139}
\affiliation{$^3$Center for Radiophysics and Space Research, Cornell
University, Ithaca, NY 14853}
\begin{abstract}
We describe progress evolving an important limit of binary orbits in
general relativity, that of a stellar mass compact object gradually
spiraling into a much larger, massive black hole.  These systems are
of great interest for gravitational wave observations.  We have
developed tools to compute for the first time the radiated fluxes of
energy and angular momentum, as well as instantaneous snapshot
waveforms, for generic geodesic orbits.  For special classes of
orbits, we compute the orbital evolution and waveforms for the
complete inspiral by imposing global conservation of energy and
angular momentum.  For fully generic orbits, inspirals and waveforms
can be obtained by augmenting our approach with a prescription for the
self force in the adiabatic limit derived by Mino.  The resulting
waveforms should be sufficiently accurate to be used in future
gravitational-wave searches.
\end{abstract}
\maketitle

The late dynamics of a merging compact binary remains one of the
greatest challenges of general relativity (GR).  GR doesn't have a
``two body'' problem so much as it has a ``one spacetime'' problem:
one must use the Einstein field equations to find the dynamical
spacetime describing a multibody system.  Although numerical
relativity has made great progress in recent years (e.g.,
{\cite{bs03}} and references therein), most astrophysically relevant
progress has come from identifying a small parameter which defines a
perturbative expansion.  One such approach is {\it post-Newtonian}
(PN) theory (e.g., {\cite{b02}} and references therein) ---
essentially, an expansion in interaction potential $GM/r c^2$.  PN
theory works very well when the bodies' separation $r$ is large.  As
the bodies come close, the expansion must be iterated to high order
(though it may be possible to modify the expansion to improve its
convergence {\cite{damour}}).

Strong field binaries can be modeled very accurately if one member is
far more massive than the other.  The spacetime is then that of the
larger body plus a perturbation due to the smaller body, with the mass
ratio acting as an expansion parameter.  This limit is not just of
formal interest: binaries consisting of ``small'' compact bodies
(white dwarfs, neutron stars, or black holes with mass $\mu \sim
1-100\,M_\odot$) captured onto highly eccentric orbits of massive
black holes ($M \sim 10^5 - 10^7\,M_\odot$) are important targets for
space-based gravitational-wave (GW) antennae, especially the LISA
{\cite{lisa}} mission.  Such captures are estimated to occur with a
rate density $\sim10^{\pm1}$ Gpc$^{-3}$ yr$^{-1}$.  Convolving with
LISA's planned sensitivity, one expects to measure dozens to thousands
of events per year {\cite{g04}}.  Precisely measuring GW phase as the
smaller body spirals into the black hole (driven by GW backreaction)
and fitting to detailed models will determine system parameters with
extraordinary accuracy.  For example, the hole's mass and spin should
be determined to $\lesssim 0.1\%$ {\cite{bc04}}.  It should even be
possible to ``map'' the black hole's spacetime, testing whether it
satisfies GR's stringent requirements {\cite{r97,ch04}}.

Thanks to their extremal mass ratio, a formal prescription for
modeling such systems now exists.  At lowest order, the small body
follows a geodesic orbit of the black hole [e.g., Ref.\ {\cite{mtw}},
Eqs.\ (33.32a)--(33.32d)].  This must be corrected by the small body's
interaction with its own spacetime distortion.  The electromagnetic
manifestation of this {\it self force} was given by Dirac
{\cite{d38}}, yielding the Abraham-Lorentz-Dirac equation of motion.
Equations for a body's curved spacetime self interaction were worked
out by Mino, Sasaki, and Tanaka {\cite{mst97}} and by Quinn and Wald
{\cite{qw97}}; see Ref.\ {\cite{p04a}} for an overview.  Many
{\cite{selffolks}} researchers are now working to develop practical
schemes to compute the self force in black hole spacetimes, which is
quite a challenge.

\par\noindent{\it Adiabatic radiation reaction (ARR)}. Astrophysical
extreme-mass-ratio binaries allow a significant simplification for
most of the inspiral, up to the last few orbits before the final
plunge and merger.  In this regime, the system evolves {\it
adiabatically}: the radiation reaction time $T_{\rm rad}$ is much
larger than the orbital time $T_{\rm orb}$: $T_{\rm orb} / T_{\rm rad}
\sim \mu/M \ll 1$.  The inspiral can be approximated as a flow through
a sequence of geodesic orbits.  To compute the leading-order,
adiabatic waveforms, it is necessary only to know the time-averaged
rates of change of the three constants of motion: the energy $E$,
axial angular momentum $L_z$, and Carter constant $Q$
{\cite{mtw,tocome}}.

This approximation can be understood by expanding the small body's
self acceleration ${\bf a}$.  Define $\varepsilon \equiv \mu/M$.  We
can then write ${\bf a} = \varepsilon \left[{\bf a}_0^{\rm diss} +
{\bf a}_0^{\rm cons} + \varepsilon \left({\bf a}_1^{\rm diss} + {\bf
a}_1^{\rm cons} \right) + \ldots\right]$.  Terms labeled ``diss''
describe {\it dissipative} aspects of the self force; they drive the
inspiral.  Those labeled ``cons'' are {\it conservative}, representing
non-dissipative components that contribute to the inspiraling body's
inertia.  Dissipative terms accumulate secularly; conservative pieces
do not.  Their observable impact can be seen in the following
expression obtained from a two-time expansion \cite{twotime,tocome}
for the azimuthal orbital phase $\Phi(t)$ (and correspondingly, the GW
phase of each harmonic component of the waveform):
\begin{equation}
\Phi(t) = \frac{1}{\varepsilon}\left[\Phi_0(t,\varepsilon t) +
  \varepsilon\Phi_1(t,\varepsilon t) + O(\varepsilon^2)\right]\;.
\label{eq:phase}
\end{equation}
The leading-order, adiabatic waveforms contain only the term $\Phi_0$
and omit the subleading term $\Phi_1$ which contributes a phase
correction of order unity over the entire inspiral.
The leading-order term is determined by ${\bf a}_0^{\rm diss}$.
Because it does not accumulate secularly, ${\bf a}_0^{\rm cons}$ does
{\it not} contribute at leading order; it (along with ${\bf a}_1^{\rm
diss}$) contributes to the subleading term $\Phi_1(t)$.

Estimates based on post-Newtonian theory suggest that the
leading-order, adiabatic waveforms will be sufficiently accurate to
{\it detect} waves in LISA data, and to provide an initial estimate of
source parameters \cite{unpub_est}.  The phase mismatch between
$\Phi_0(t)$ and the true signal is compensated for by adjustments in
model parameters, which introduces systematic error in the inferred
parameters.  Eliminating this systematic error will require
``measurement templates'' that accurately model $\Phi_1(t)$ --- a far
more difficult task.

In this paper, we describe recent progress in computing adiabatic
waveforms, using (i) an approach called ``Poor man's radiation
reaction'' (PMRR) for special classes of orbits, and (ii) augmenting
PMRR using recent results of Mino {\cite{m03}} for fully generic
orbits.

\smallskip
\par\noindent{\it Poor man's radiation reaction.}  PMRR only requires
knowledge of the energy and angular momentum the binary radiates;
imposing global conservation, we evolve those quantities,
approximating inspiral as a flow through a sequence of orbits.  PMRR
cannot rigorously evolve {\it all} the constants $E$, $L_z$ and $Q$
describing black hole orbits, except in special cases.  It is simple
to evolve $E$ and $L_z$ --- one computes the rate at which $E$ and
$L_z$ are carried to infinity {\cite{i68}} and absorbed by the hole's
event horizon {\cite{hh72}}, and imposes global conservation
{\cite{qw99}}: $\dot E^{\rm orb} + \dot E^{\rm rad} = 0$, $\dot
L_z^{\rm orb} + \dot L_z^{\rm rad} = 0$.  Carter's constant $Q$ cannot
be so evolved, since there is no notion of $Q$-flux carried by
radiation and no associated global conservation law.  Orbits with zero
eccentricity or inclination are sufficiently constrained that $\dot Q$
is fixed by $\dot E$ and $\dot L_z$ {\cite{circtocirc}}.

The general case admits no such constraints.  Nevertheless we can
produce inspirals accurate enough for data-analysis algorithm
development using simple, crude approximations based on limiting
cases. For example, forcing a certain inclination angle to be constant
determines the inspiral using only $E$ and $L_z$ fluxes
{\cite{ghk02}}.  Such inspirals are very unlikely to be accurate
enough for detection templates.

The central engine of PMRR is a complex function $\psi_4$ representing
the small body's perturbation to the black hole's curvature.  This
function is the Weyl (vacuum) curvature tensor $C_{abcd}$ projected
onto a set of null vectors that are very convenient for describing
outgoing radiation.  Far away, $\psi_4$ is simply related to the two
GW polarizations:
\begin{equation}
\psi_4(r\to\infty) = \frac{1}{2}\frac{\partial^2}{\partial t^2}
\left(h_+ - i h_\times\right)\;.
\label{eq:psi4_big_r}
\end{equation}
From this one finds the outgoing fluxes of $E$ and $L_z$ {\cite{i68}}.
This $\psi_4$ also completely describes the radiation's interaction
with the black hole, and thus encodes the rate at which the large
black hole absorbs $E$ and $L_z$ {\cite{hh72,tp74}}.

Teukolsky derived a ``master equation'' for black hole perturbations
{\cite{t73}}, and showed it separates by expanding in Fourier modes
and spheroidal harmonics:
\begin{equation}
\psi_4 = \frac{1}{(r - ia\cos\theta)^4}\int d\omega \sum_{lm}
R_{lm\omega}(r)S_{lm\omega}(\theta)e^{i(m\phi - \omega t)}\;.
\label{eq:modal_exp}
\end{equation}
(The parameter $a = |\vec S|/M$ is the black hole spin per unit mass.)
It is not necessary to expand in modes: one can leave the Teukolsky
equation as a PDE coupled in $t$, $r$, and $\theta$ (the $\phi$
dependence trivially decouples), and evolve initial data for
$\psi_4$. This works so well modeling source-free radiation
{\cite{timedomain}} that it has been argued time domain methods may
replace frequency decomposition for most applications {\cite{p04b}}.
For point particle sources, frequency domain methods are currently
considerably more accurate {\cite{td}}.

The radial functions $R_{lm\omega}(r)$ are obtained by solving
\begin{equation}
\Delta^2\left(\Delta^{-1} R'_{lm\omega}\right)' - V_{lm\omega}(r)
R_{lm\omega} = -{\cal T}_{lm\omega}(r)\;,
\label{eq:teuk}
\end{equation}
where prime denotes $d/dr$, $\Delta = r^2 - 2 M r + a^2$,
$V_{lm\omega}$ is a potential [e.g., {\cite{h00}}, Eq.\ (4.3)], and
${\cal T}_{lm\omega}$ is a source constructed from the small body's
stress energy tensor.  We build a Green's function
$G_{lm\omega}(r,r')$ from solutions to the homogeneous wave equation
(setting ${\cal T}_{lm\omega} = 0$), using certain analytic
transformations {\cite{sn82}} that allow high numerical accuracy.  We
then find $R_{lm\omega}(r)$ by integrating $G_{lm\omega}(r,r')$ over
the source,
\begin{equation}
R_{lm\omega}(r) = -\int dr' G_{lm\omega}(r,r'){\cal
T}_{lm\omega}(r')\;.
\label{eq:formal_soln}
\end{equation}
Our particular interest is in the limits $r\to\infty$ and $r\to$ the
event horizon; from the function in these limits, we extract the $E$
and $L_z$ fluxes mode-by-mode. Using many modes, we assemble the
adiabatic rates of change $\langle\dot E\rangle$ and $\langle\dot
L_z\rangle$.

\begin{figure}[ht]
\includegraphics[width = .45\textwidth]{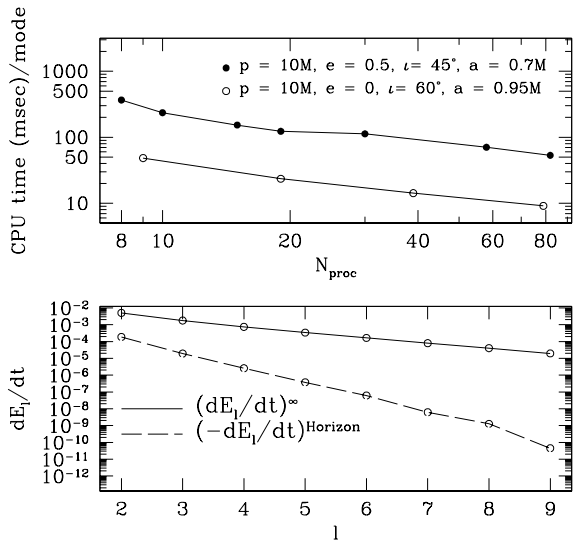}
\vskip -0.3cm
\caption{Top: Average CPU time per mode versus number of processors
$N$ for two strong field orbits [circular ($e = 0$) and eccentric ($e
= 0.5$)].  Our scaling shows $T_{\rm cpu} \propto 1/N$, showing how
well PMRR parallelizes.  The typical CPU time per mode is fairly
short, though eccentric orbits are about 5--10 times slower than
circular.  Bottom: Flux $dE_l/dt$ (defined in the text) versus $l$ for
an orbit with $p = 4$, $e = 0.5$, and $\iota = 45^\circ$; the black
hole's spin $a = 0.9M$.  We generically find that the flux sums
converge exponentially with $l$.}
\vskip -0.3cm
\label{fig1}
\end{figure}

For bound orbits, the source is nicely described by a harmonic
expansion.  The continuous frequency $\omega$ goes over to a discrete
set $\omega_{mkn} = m\Omega_\phi + k\Omega_\theta + n\Omega_r$, where
$\Omega_{\phi,\theta,r}$ describe motion in $\phi$, $\theta$, and $r$
{\cite{s02,dh04}}.  Our modes become 4 index objects, $(lmkn)$.  There
is no mode-mode coupling, so this problem is ideally suited to
parallel computation: different modes can be sent to different
processors with little communication cost.  Figure {\ref{fig1}}
illustrates parallelization.  The top panel shows the average CPU time
per mode $(lmkn)$ as a function of number of processors $N$.  We find
nearly linear scaling in $1/N$, demonstrating that we have little
parallelization overhead.  Typical CPU time per mode is quite short,
so that $\sim 10^4-10^5$ modes can be computed in a reasonable time.

The bottom panel shows the convergence of energy flux from a strong
field orbit.  We write the orbit's radial motion as $r = p M/(1 +
e\cos\psi)$; for this plot, we put eccentricity $e = 0.5$ and
semi-latus rectum $p = 4$.  The inclination $\iota = 45^\circ$; the
black hole's spin is $a = 0.9M$.  (Precise definitions of $p$, $e$,
and $\iota$ can be found in {\cite{dh04}}; their key property for this
paper is that they are simply related to the constants $E$, $L_z$, and
$Q$.)  We plot
\begin{equation}
\dot E_l = \sum_{m = -l}^l\sum_{k = -K}^K\sum_{n = -N}^N \dot E_{lmkn}
\label{eq:edot_l}
\end{equation}
where $\dot E_{lmkn}$ is the modal energy flux.  The cutoff values $N$
and $K$ (which formally are $\infty$), are set large enough that
neglected terms contribute a fractional amount less than $10^{-4}$ to
the sum; details of this truncation will be presented elsewhere
{\cite{dh_inprep}}.  We generically find that $\dot E_l$ falls
exponentially with $l$.  The orbital parameter's influence on the rate
of this falloff, and on other convergence criteria, will be presented
in {\cite{dh_inprep}}.  Most importantly, the convergence of these
fluxes is fairly quick for $e \lesssim 0.7$, a very astrophysically
important range {\cite{g04}}.

\smallskip
\par\noindent{\it Mino's adiabatic self force and $Q$'s evolution.}
Recent work by Mino {\cite{m03}} makes it possible to compute the time
average $\langle \dot Q \rangle$ of ${\dot Q}$, allowing us to compute
$\Phi_0(t)$ in Eq.\ ({\ref{eq:phase}) for generic orbits.  The
regularized self force $f^b$ involves an integral over the orbiting
body's past worldline {\cite{mst97,qw97}}.  Mino shows that this
integral can be replaced in the adiabatic limit by a relatively simple
expression involving the difference between ``retarded'' and
``advanced'' forces: $ f^b = (f^b_{\rm ret} - f^b_{\rm adv})/2$.  The
``retarded'' force depends on events on the orbiting body's past
lightcone; the ``advanced'' force depends on the future lightcone.
Their difference removes the divergent point-particle self
interaction; the remaining force drives the inspiral.  This result is
strikingly similar to Dirac's result {\cite{d38}}, and indeed
reproduces the rule posited (without proof) by Gal'tsov {\cite{g82}}.

The Carter constant is given by $Q = Q_{ab} p^a p^b$, where $Q_{ab}$
is a Killing tensor and $p^a$ is the small body's 4-momentum.  Taking
a time derivative yields ${\dot Q} = 2 Q_{ab} p^a {\dot p}^b = 2
Q_{ab} p^a f^b$.  We express the self force in terms of the radiative
Green's function using Mino's result, and use the expansion of the
radiative Green's function in terms of modes {\cite{g82}}.  Time
averaging yields an expression of the form {\cite{tocome}}
\begin{equation}
\langle {\dot Q} \rangle = \sum_{r=\infty,H} \sum_{lmkn}
W[{\cal Q}_{lmkn}(r), R_{lmkn}(r)].
\label{eq:Qdot2}
\end{equation}
The quantities $R_{lmkn}(r)$ come from $R_{lm\omega}(r)$ by writing
$R_{lm\omega}(r) = \sum_{kn} R_{lmkn}(r) \delta(\omega -
\omega_{mkn})$.  The first sum in Eq.\ (\ref{eq:Qdot2}) means that the
expression is evaluated near the horizon and near infinity.  The
quantities ${\cal Q}_{lmkn}(r)$ are computed via an integral similar
to that in Eq.\ (\ref{eq:formal_soln}), but with the source ${\cal
T}_{lm\omega}$ replaced by a new source built from the Killing tensor
$Q_{ab}$ and the orbiting body's 4-velocity, and evaluated at $\omega
= \omega_{mkn}$.  Finally $W$ is the Wronskian which is independent of
$r$ near the horizon and near infinity as both $R_{lmkn}(r)$ and
${\cal Q}_{lmkn}(r)$ satisfy the homogeneous Teukolsky equation at
those locations.  Details of this calculation will be presented
elsewhere {\cite{tocome}}.  Using this result it will be as
straightforward to compute $\langle {\dot Q} \rangle$ as it is
currently to compute $\langle {\dot E} \rangle$ and $\langle {\dot
L}_z \rangle$.

\par\noindent{\it Applications and future directions.}  Building
inspiral waveforms requires us to compute backreaction effects upon a
dense ``grid'' of orbits in parameter space.  Each orbit is
represented by a coordinate $(p,e,\iota)$; backreaction gives the
tangent $(\dot p, \dot e, \dot\iota)$ to the inspiral trajectory.  We
flow along this tangent vector, building the trajectory
$[p(t),e(t),\iota(t)]$.  Choosing initial conditions, we likewise
build the inspiral worldline $z^a(t) = [t, r(t), \theta(t), \phi(t)]$.
To date, only circular ($e = 0$) {\cite{h01}} and equatorial ($\iota =
0,180^\circ$) {\cite{gk02}} inspirals have been computed; the general
case is under development {\cite{dh_inprep}}, though we can present
waveform ``snapshots'' (Fig.\ {\ref{fig2}}).

\begin{figure}[ht]
\includegraphics[width = .45\textwidth]{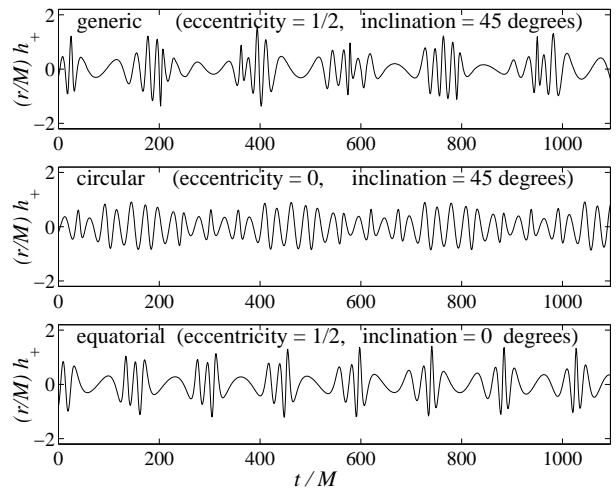}
\caption{Gravitational waveform for several strong field orbits; all
are for a hole with spin $a = 0.9M$, and have $p = 4$.  Top panel:
eccentricity $e = 0.5$, inclination $\iota = 45^\circ$.  Middle: $e =
0$, $\iota = 45^\circ$.  Bottom: $e = 0.5$, $\iota = 0^\circ$.  The
orbital parameters, particularly eccentricity, richly influence the
wave's harmonic structure.}
\label{fig2}
\vskip -0.2cm
\end{figure}

The cases we have studied in detail demonstrate the rich physics
encoded by these events.  Three particularly interesting features are:

\par\noindent $\bullet$ {\it Tidal coupling to the event horizon}: The
orbiting body raises a tidal bulge on the black hole.  This bulge
interacts with the orbit, transferring the hole's spin to the orbit,
just as planetary tides can transfer angular momentum to a satellite.
If the hole rotates rapidly and the orbit has shallow inclination, the
tide can significantly prolong the inspiral {\cite{h01}}.

\par\noindent $\bullet$ {\it Strong field precessions}: The time to
oscillate through $r$ does not equal the time to move through $2\pi$
radians of $\phi$.  The mismatch between these timescales gives
perihelion precession, a classic GR test.  With black holes, this
effect can amount to thousands of radians per orbit --- the small body
``whirls'' many times near the hole before ``zooming'' out to large
radius.  This ``zoom-whirl'' behavior leaves a distinctive stamp on
the waveform {\cite{gk02}, seen in the short, high-frequency segments
in Fig.\ {\ref{fig2}}.

\par\noindent $\bullet$ {\it Spin-orbit coupling}: The black hole's
spin makes the spacetime geometry oblate, and drags spacetime into
co-rotation with it.  This splits the $\phi$ and $\theta$ frequencies,
introducing further modulations to the waveform (middle panel of Fig.\
{\ref{fig2}}).

All three of these features richly influence the phasing of a binary's
GWs, and so are readily discernible in a phase coherent measurement.
It is through determination of these features and their time
evolutions that the binary's parameters and its strong field spacetime
structure can be determined so well.

Future work will develop ARR and waveforms with a focus upon templates
for future GW searches.  One goal is to use spectral methods for
solving many of the formalism's equations.  Such methods are typically
exponentially convergent in the number of basis functions.  Since ARR
requires many multipoles, each must be as accurate as possible.  We
are very encouraged by the work of Fujita and Tagoshi {\cite{ft04}},
who demonstrate that a particular set of basis functions allows modal
solutions with essentially double precision accuracy.

\noindent{\it Acknowledgments.}  This work was supported at MIT by NSF
Grant PHY-0244424 and NASA Grant NAGW-12906, and at Cornell by NSF
Grant PHY-0140209 and NASA/New York Space Grant Consortium.  Most
computations were performed using the MIT Astrophysics Division
Beowulf cluster; we thank Edmund Bertschinger, Omri Schwarz, and Alex
Shirokov for cluster assistance.  We thank Saul Teukolsky for helpful
conversations.


\begin{thebibliography}{99}

\bibitem{bs03} T.\ W.\ Baumgarte and S.\ L.\ Shapiro, Phys.\ Rept.\
  {\bf 376}, 41 (2003).

\bibitem{b02} L.\ Blanchet, Living Rev.\ Relativity {\bf 5}, 3 (2002).

\bibitem{damour} A.\ Buonanno and T.\ Damour, Phys.\ Rev.\ D {\bf 59},
  084006 (1999); T.\ Damour, {\it ibid.}\ {\bf 64}, 124013 (2001); T.\
  Damour, B.\ R.\ Iyer, P.\ Jaranowski, and B.\ S.\ Sathyaprakash,
  {\it ibid.}\ {\bf 67}, 064028 (2003).

\bibitem{lisa} http://lisa.nasa.gov

\bibitem{g04} J.\ R.\ Gair {\it et al.}, Class.\ Quantum Grav.\ {\bf
  21}, S1595 (2004).

\bibitem{bc04} L.\ Barack and C.\ Cutler, Phys.\ Rev.\ D {\bf 69},
  082005 (2004).

\bibitem{r97} F.\ Ryan, Phys.\ Rev.\ D {\bf 56}, 1845 (1997).

\bibitem{ch04} N.\ A.\ Collins and S.\ A.\ Hughes, Phys.\ Rev.\ D {\bf
  69}, 124022 (2004).

\bibitem{mtw} C.\ W.\ Misner, K.\ S.\ Thorne, and J.\ A.\ Wheeler,
  {\it Gravitation} (Freeman, San Francisco, 1973).

\bibitem{d38} P.\ A.\ M.\ Dirac, Proc.\ R.\ Soc.\ London, Ser.\ A {\bf
  167}, 148 (1938).

\bibitem{mst97} Y.\ Mino, M.\ Sasaki, and T.\ Tanaka, Phys.\ Rev.\ D
  {\bf 55}, 3457 (1997).

\bibitem{qw97} T.\ Quinn and R.\ M.\ Wald, Phys.\ Rev.\ D {\bf 56},
  3381 (1997).

\bibitem{p04a} E.\ Poisson, Living Rev.\ Relativity, {\bf 7}, 6
  (2004).

\bibitem{selffolks} L.\ Barack and C.\ O.\ Lousto, Phys.\ Rev.\ D {\bf
  66}, 061502 (2002); L.\ Barack and A.\ Ori, Phys.\ Rev.\ Lett.\ {\bf
  90}, 111101 (2003); S.\ Detweiler and B.\ F.\ Whiting, Phys.\ Rev.\
  D {\bf 67}, 024025 (2003); W.\ Hikida {\it et al.}, Prog.\ Theor.\
  Phys.\ {\bf 111}, 821 (2004); S.\ Detweiler and E.\ Poisson, Phys.\
  Rev.\ Lett.\ {\bf 69}, 084019 (2004).

\bibitem{twotime} J.\ Kevorkian and J.\ D.\ Kole, {\it Multiple scale
  and singular perturbation methods} (Springer, New York, 1996),
  Sec.\ 3.6.

\bibitem{tocome} S.\ Drasco, \'E.\ \'E.\ Flanagan and S.\ A.\ Hughes,
  in preparation.

\bibitem{unpub_est} For example, for a $10 M_\odot$ circular inspiral
  into a $10^6 M_\odot$, $a=0.999M$ black hole, the $\mu/M$
  corrections in the post-3.5-Newtonian phase of the Fourier transform
  of the waveform that correspond to $\Phi_1(t)$ give a phase error
  between $0.003$ Hz and $0.03$ Hz (the last year of inspiral
  \cite{finnthorne}) of $\le 0.8$ cycles (after optimizing
  time-of-arrival and overall phase).  For detection purposes, one
  needs phase coherence for only $\sim 3$ weeks \protect{\cite{g04}};
  the phase error over this timescale will be even smaller.

\bibitem{finnthorne} L.\ S.\ Finn and K.\ S.\ Thorne, Phys.\ Rev.\ D
  {\bf 62}, 124021 (2000).

\bibitem{m03} Y.\ Mino, Phys.\ Rev.\ D {\bf 67}, 084027 (2003).

\bibitem{i68} R.\ Isaacson, Phys.\ Rev.\ {\bf 166}, 1272 (1968).

\bibitem{hh72} S.\ W.\ Hawking and J.\ B.\ Hartle, Commun.\ Math.\
  Phys.\ {\bf 25}, 283 (1972).

\bibitem{qw99} T.\ Quinn and R.\ M.\ Wald, Phys.\ Rev.\ D {\bf 60},
  064009 (1999)

\bibitem{circtocirc} F.\ D.\ Ryan, Phys.\ Rev.\ D {\bf 53}, 3064
  (1996); D.\ Kennefick and A.\ Ori, {\it ibid.}\ {\bf 53}, 4319
  (1996); Y.\ Mino, M.\ Sasaki, M.\ Shibata, H.\ Tagoshi, and T.\
  Tanaka, Prog.\ Theor.\ Phys.\ Suppl.\ {\bf 128}, 1 (1997).

\bibitem{ghk02} K.\ Glampedakis, S.\ A.\ Hughes, and D.\ Kennefick,
  Phys.\ Rev.\ D {\bf 66}, 064005 (2002).

\bibitem{tp74} S.\ A.\ Teukolsky and W.\ H.\ Press, Astrophys.\ J.\
  {\bf 193}, 443 (1974).

\bibitem{t73} S.\ A.\ Teukolsky, Astrophys.\ J.\ {\bf 185}, 635
  (1973).

\bibitem{timedomain} W.\ Krivan, P.\ Laguna, and P.\ Papadopoulos,
  Phys.\ Rev.\ D {\bf 54}, 4728 (1997); W.\ Krivan, P.\ Laguna, P.\
  Papadopoulos, and N.\ Andersson, {\it ibid.}\ {\bf 56}, 3395 (1997);
  L.\ M.\ Burko and G.\ Khanna, {\it ibid.}\ {\bf 67}, 081502(R)
  (2004); M.\ A.\ Scheel {\it et al.}, {\it ibid.}\ {\bf 69}, 104006
  (2004); K.\ Martel, unpublished Ph.\ D.\ thesis, University of
  Guelph (2003).

\bibitem{p04b} E.\ Poisson, Phys.\ Rev.\ D {\bf 70}, 084044 (2004).

\bibitem{td} It may be possible to combine a frequency-domain inspiral
  calculation with a time-domain approach to the emitted radiation.
  The averaging needed for ARR is naturally done in the frequency
  domain; time domain codes are superior for radiation generation and
  propagation.  A hybrid approach could combine the best features of
  both toolkits.

\bibitem{h00} S.\ A.\ Hughes, Phys.\ Rev.\ D {\bf 61}, 084004 (2000).

\bibitem{sn82} M.\ Sasaki and T.\ Nakamura, Prog.\ Theor.\ Phys.\ {\bf
  67}, 1788 (1982).

\bibitem{s02} W.\ Schmidt, Class.\ Quantum Grav.\ {\bf 19}, 2743
  (2002).

\bibitem{dh04} S.\ Drasco and S.\ A.\ Hughes, Phys.\ Rev.\ D {\bf 69},
  044015 (2004).

\bibitem{dh_inprep} S.\ Drasco and S.\ A.\ Hughes, in preparation.

\bibitem{g82} D.\ V.\ Gal'tsov, J.\ Phys.\ A {\bf 15}, 3737 (1982);
  particularly Eq.\ (4.4) and associated discussion.

\bibitem{h01} S.\ A.\ Hughes, Phys.\ Rev.\ D {\bf 64}, 064004 (2001).

\bibitem{gk02} K.\ Glampedakis and D.\ Kennefick, Phys.\ Rev.\ D {\bf
  66}, 044002 (2002).

\bibitem{ft04} R.\ Fujita and H.\ Tagoshi, Prog.\ Theor.\ Phys.\ {\bf
  112}, 415 (2004).

\end{thebibliography}
\end{document}